\documentclass[prb, twocolumn, aps, superscriptaddress]{revtex4}
\usepackage{graphicx}
\usepackage{bm}
\usepackage{amssymb}
\usepackage{amsmath}
\usepackage{amsfonts}
\newcommand {\ket} [1] {| #1 \rangle}
\newcommand {\bkt} [1] {\langle #1 \rangle}
\newcommand {\dbkt} [2] {\langle #1 | #2 \rangle}
\newcommand {\tbkt} [3] {\langle #1 | #2 | #3 \rangle}
\newcommand {\pd} [2] {\frac{\partial #1}{\partial #2}}
\newcommand {\td} [2] {\frac{d #1}{d #2}}

\begin{document}
\title{Current-induced spin torques in III-V ferromagnetic semiconductors}
\author{Dimitrie Culcer}
\affiliation{Advanced Photon Source, Argonne National Laboratory,
Argonne, IL 60439.} \affiliation{Northern Illinois University, De
Kalb, IL 60115.}
\author{M.~E.~Lucassen}
\author{R.~A.~Duine}
\affiliation{Institute for Theoretical Physics, Utrecht
University, Leuvenlaan 4, 3584 CE, Utrecht, The Netherlands}
\author{R.~Winkler}
\affiliation{Advanced Photon Source, Argonne National Laboratory,
Argonne, IL 60439.} \affiliation{Northern Illinois University, De
Kalb, IL 60115.}

\begin{abstract}
  We formulate a theory of current-induced spin torques in
  inhomogeneous III-V ferromagnetic semiconductors. The carrier
  spin-3/2 and large spin-orbit interaction, leading to spin
  non-conservation, introduce significant conceptual differences
  from spin torques in ferromagnetic metals. We determine the spin
  density in an electric field in the weak momentum scattering
  regime, demonstrating that the torque on the magnetization is
  intimately related to spin precession under the action of both the
  spin-orbit interaction and the exchange field characteristic of
  ferromagnetism. The spin polarization excited by the electric
  field is smaller than in ferromagnetic metals and, due to lack of
  angular momentum conservation, cannot be expressed in a simple
  closed vectorial form. Remarkably, scalar and spin-dependent
  scattering do not affect the result. We use our results to
  estimate the velocity of current-driven domain walls.
\end{abstract}
\date{\today}
\maketitle

\section{Introduction}

Over a decade ago, Slonczewski \cite{slo96} and Berger \cite{ber96}
predicted that an electrical current induces a torque on the
magnetization of a ferromagnetic metal, and subsequent research has
since identified distinct contributions called the reactive spin
transfer torque and the dissipative spin transfer torque, sometimes
referred to as adiabatic and non-adiabatic, respectively.
\cite{ral07} Progress on the understanding of this effect in metals
has been steady, \cite{baz98, tsoi98, sti02, ans04, li04, thi05,
bar05, dui07, koh06, tse06, pie07} and the field has been stimulated
by applications in spintronics and nanotechnology, as a way to
manipulate magnetization and thus information.

Current-induced spin torques arise from a small mismatch between the
spin polarization of conduction electrons and the magnetization
present throughout a material, and reflect the nonlocal nature of
magnetization dynamics in inhomogeneous systems. \cite{tse05} They
are the converse of processes such as giant magnetoresistance (GMR).
The calculation of spin torques is equivalent to finding the
conduction-electron spin density in an electric field, and in metals
they can be easily expressed as the divergence of the spin current.
\cite{tse06} Recently spin torque-related effects have been
investigated in metal-based systems and nanomagnets, e.g.,
spin-torque driven ferromagnetic resonance, \cite{kup06} spin
torques in nanomagnets \cite{wai05} and in continuously variable
magnetizations, \cite{xiao06} the $s$-$d$ interaction in
inhomogeneous ferromagnets, \cite{reb05} current-induced
magnetization dynamics, \cite{fer04} and domain wall motion.
\cite{tat06}

Despite these efforts, spin torques have been little studied in
materials displaying some of the richest physics and technological
promise, namely ferromagnetic semiconductors. The most intensely
studied ferromagnetic semiconductors are Mn-doped III-V compounds,
which fall into two classes: zincblende lattices, such as (Ga,Mn)As
and wurtzite lattices, such as (Ga,Mn)N. The former have been much
investigated \cite{jun06, koen03} and will be the subject of this
work, while the latter are awaiting further experimental
developments. \cite{kea05} Recent experiments have succeeded in
fabricating (Ga,Mn)As spin transfer torque devices, \cite{wun07}
while domain wall motion \cite{yamanouchi06,nguy07} and resistance
\cite{nguy06} have been addressed, and spin transfer physics was the
subject of a recent review. \cite{die07}

Ferromagnetic semiconductors differ profoundly from ferromagnetic
metals. Being $p$-type, the carriers are holes described by an
effective spin-3/2 and subject to a strong spin-orbit interaction.
The hole spin is thus not conserved. The band Hamiltonian is the
Luttinger Hamiltonian, \cite{lut56} which, in the spherical
approximation, is composed of a series of multipoles in spin space.
\cite{lip70} The hole spin is a dipole, while the spin-orbit
interaction is a quadrupole, and the precession of the spin in a
quadrupole field is highly nontrivial. \cite{dim06} Magnetism is due
to localized Mn moments, with the exchange interaction mediated by
itinerant holes. In the mean-field approximation this interaction is
taken into account as an effective field (a dipole) acting on the
carriers and produces a splitting comparable in magnitude to the
Fermi energy. \cite{jun06} The combined effect of this dipole
exchange field and the quadrupole spin-orbit interaction is not a
simple additive problem. \cite{dim06} These materials are often in
the weak momentum scattering limit, \cite{jun06} i.e.,
$\varepsilon_F \tau/\hbar \gg 1$, where $\varepsilon_F$ is the Fermi
energy and $\tau$ is a characteristic scattering time. Due to the
fast spin precession as a result of the spin-orbit interaction the
relaxation time approximation is inappropriate to describe spin
dynamics. The carrier spin is not conserved which introduces
arbitrariness in the definition of the spin current. \cite{shi06}
Therefore, spin torques cannot be identified, as in metallic
systems, \cite{tse06} with the divergence of the spin current.
Finally electrically-induced spin densities \cite{ivc78} in bulk
nonmagnetic zincblende semiconductors are forbidden by symmetry,
raising the question of what form the spin density may take in such
a system when the symmetry is lowered by the magnetization.

In this paper we present a general theory of current-induced spin
torques in zincblende III-V ferromagnetic semiconductors, which, to
our knowledge, is the first of its kind. Our study is based on the
Luttinger Hamiltonian \cite{lut56} and it assumes $\varepsilon_F
\tau/\hbar \gg 1$, a time-independent magnetization with small
spatial gradients, zero temperature, no compensation and
short-ranged impurity potentials (justified by the large carrier
densities in ferromagnetic semiconductors and the short-range nature
of the exchange interaction \cite{jun06}). We take the spin-orbit
interaction to be the \emph{dominant} term and treat the exchange
field in first-order perturbation theory. This is a good
approximation for the lower end of Mn concentrations ($\approx
2-5\%$). We neglect also quantum interference effects such as weak
localization.

In recent years a number of microscopic theories of spin torques in
ferromagnetic metals have been developed. \cite{dui07, pie07, koh06,
tse06} A common strategy is to begin with a uniformly magnetized
state and consider small perturbations around it. The method used in
this work is slightly different, although the physics is the same
and no difference is expected in the final results. We assume a
magnetization that is a function of position and has small
gradients, and we work to first order in the gradient of the
magnetization. We would like to note that a recent study has been
published which considers spin torques in metals starting from a
spin continuity equation. \cite{Tatara}

The outline of this paper is as follows. The next section contains a
derivation of the kinetic equation that will be used in the
remainder of the paper. We begin from the quantum Liouville equation
and project it in momentum space, then introduce so-called Wigner
coordinates and derive the equation satisfied by the Wigner
distribution. In Section III this equation is solved in the presence
of an electric field, and in section IV the spin density induced by
the electric field is found, which gives the spin torque acting on
the magnetization. The form and implications of the results are
discussed, and their applicability to GaMnAs is demonstrated.
Finally, the domain wall velocity as a result of the spin torque is
estimated in Sec.~\ref{sec:dws}.

\section{Kinetic equation}

The typical setup for a spin torque experiment consists of two slabs
of ferromagnetic material with non-collinear magnetizations,
separated by a tunnel barrier. Since the magnetizations of the two
slabs are non-collinear there is a region near the interface over
which the magnetization changes. To determine the continuum limit of
this setup one can begin by visualizing a large number of slabs put
together, with slight variations in the direction of the
magnetization of each slab. Then one can imagine the interfaces
between the slabs disappearing, leaving one large sample with an
inhomogeneous magnetization, in such a way that the gradient of the
magnetization varies little over distances comparable to the lattice
spacing. The gradient expansion in the magnetization that follows
from this procedure is valid as long as the length scale on which
the magnetization varies is much longer than the relevant length
scales of the carriers, i.e., the Fermi wavelength and mean free
path.

Spin torques appear when an electrical current flows through such a
material. In a ferromagnetic semiconductor the magnetization is a
result of the Mn ions which interact by means of the exchange
coupling mediated by itinerant holes. The holes themselves have a
spin polarization, and in equilibrium the hole spin polarization
follows the magnetization. When an electrical current flows through
the sample one can think for example of a hole which is taken from
position ${\bm r}$, where its spin is parallel to the local
magnetization at ${\bm r}$, and transporting it to ${\bm r} +
\delta{\bm r}$. The magnetization at ${\bm r} + \delta{\bm r}$ is
slightly different from the magnetization at ${\bm r}$, so a torque
is exerted by the itinerant hole on the magnetization. What makes
the situation in ferromagnetic semiconductors more difficult and
more interesting is that while the hole is moving from ${\bm r}$ to
${\bm r} + \delta{\bm r}$ it is subject to the strong spin-orbit
interaction, which acts to randomize its spin. In order to determine
the effect of spin-orbit interactions, which are wave
vector-dependent, on the itinerant holes and ultimately on the
inhomogeneous magnetization we need to study the kinetic equation,
which takes into account both the momentum dependence and the
position dependence.

In this section we will derive a kinetic equation suitable for
describing inhomogeneous ferromagnetic semiconductors in an electric
field. We consider the system to be described by a density operator
$\hat \rho$, which obeys the quantum Liouville equation
\begin{equation}
\td{\hat\rho}{t} + \frac{i}{\hbar} \, [\hat H, \hat \rho] = 0.
\end{equation}
The total Hamiltonian $\hat H$ contains contributions due to the
band Hamiltonian $\hat H_v$, the scalar impurity potential, the
exchange interaction between delocalized holes and localized Mn
moments, and the electric field. These will be given below. The
Liouville equation is projected onto a set of states $\ket{u_{{\bm
k}s}}$ of definite wave vector ${\bm k}$ and spin $s$, which are
assumed to be Bloch functions and eigenstates of the Luttinger
Hamiltonian $\hat H_v$. The matrix elements of $\hat\rho$ in this
basis are $\rho_{ss'}({\bm k},{\bm k}') \equiv \rho({\bm k}, {\bm
k}')$ and are treated as matrices in spin space. $\hat H_v$ is
diagonal in ${\bm k}$ and its matrix elements in this basis are $H_v
\equiv H_v({\bm k})$ \cite{lut56, lip70}
\begin{equation}\label{eq:HLutt}
  H_v = \frac{\hbar^2}{2m} \left[
    \gamma_1 k^2 + \bar\gamma\left({\textstyle\frac{5}{2}}k^2 -
      2({\bm k}\cdot{\bm S})^2 \right) \right],
\end{equation}
where ${\bm S}$ is a vector of spin-3/2 matrices. The term
proportional to the Luttinger parameter $\gamma_1$ gives the hole
kinetic energy. For $k\ne 0$, the term proportional to $\bar\gamma$
separates the heavy hole (HH) and light hole (LH) states, i.e., it
is the spin-orbit coupling that plays a central role in the present
analysis. The Mn$^{2+}$ ions give rise both to a net magnetic
moment, through the hole-mediated exchange interaction, and to
scattering, which has a scalar part and a spin-dependent part. These
are contained in the Hamiltonian $H_\mathrm{Mn}$,
\begin{equation}
{H}_\mathrm{Mn} ({\bm r}) = \sum_I \big[ \mathcal{U}({\bm r} -
{\bm R}_I)\, \openone + \mathcal{V}({\bm r} - {\bm R}_I)\, {\bm
s}_I\cdot{\bm S} \big],
\end{equation}
where the sum runs over the positions ${\bm R}_I$ of the Mn$^{2+}$
ions, with ${\bm s}_I$ the Mn spin. We approximate the interactions
represented by $H_\mathrm{Mn}$ as short-ranged, so that
$\mathcal{U}({\bm r} - {\bm R}_I) = U \, \delta({\bm r} - {\bm
R}_I)$ and $\mathcal{V}({\bm r} - {\bm R}_I) = (J_{pd}/V) \,
\delta({\bm r} - {\bm R}_I)$, with $J_{pd}$ the exchange constant
between the localized Mn moments and the itinerant holes, and $V$
the sample volume. The matrix elements of $H_\mathrm{Mn}$ in the
basis $\{\ket{u_{\bm k}s}\}$ are decomposed into a part $H_{pd}$
diagonal in ${\bm k}$, which gives the net magnetization $\bm{M}$,
\begin{equation}
H_\mathrm{Mn}^{{\bm k} = {\bm k}'} \equiv H_{pd} =
\frac{N_\mathrm{Mn}\, J_{pd}}{V} \, {\bm s}\cdot{\bm S} \equiv
{\bm M} \cdot {\bm S},
\end{equation}
where $N_\mathrm{Mn}$ is the number of Mn$^{2+}$ ions and we assume
all Mn spins polarized setting ${\bm s}_I = {\bm s}$, and a part
off-diagonal in ${\bm k}$, which causes spin-dependent scattering
and will be given below. We will concentrate in this work on the
case in which the exchange splitting is smaller than the spin-orbit
coupling at the Fermi energy, i.e., $|\bm{M}| \ll 2\bar{\gamma}
\hbar^2k_F^2 / m$, where $k_F$ is the Fermi wave vector, and we work
to first order in $m |\bm M|/(2\bar{\gamma} \hbar^2k_F^2)$.

In studying inhomogeneous magnetization one must account for
real-space as well as wave-vector dependence, which is accomplished
by defining a Wigner distribution. The Wigner function corresponding
to the one-particle density matrix $\rho({\bm k}, {\bm k}')$ is
\begin{equation}
\label{eq:wigin} f_{\bf q} ({\bf r}) = \int \!\!
\frac{d^3Q}{(2\pi)^3} \, e^{i{\bf Q}\cdot{\bf r}} \rho ({\bm q} +
{\bm Q}/2, {\bm q} - {\bm Q}/2).
\end{equation}
where ${\bm q} = ({\bm k} + {\bm k}')/2$ and ${\bm Q} = {\bm k} -
{\bm k}'$. The next step is to derive an equation describing the
time evolution of the Wigner distribution $f_{\bm q} ({\bm r})$. The
kinetic equation for the Wigner function $f \equiv f_{\bm q}({\bm
r})$ is obtained by projecting the quantum Liouville equation onto
the basis states $\ket{u_{{\bm k}s}}$, then making the
transformation (\ref{eq:wigin}). The Hamiltonian $\hat H$ is
diagonal in wave vector. The first step, in which no approximations
have been made, gives us the Liouville equation in terms of the
so-called Wigner coordinates ${\bm q}$ and ${\bm Q}$
\begin{widetext}
\begin{equation}\label{frho}
\arraycolsep 0.3ex
\begin{array}{rl}
\displaystyle \pd{f_{\bm q}}{t} + \frac{i}{\hbar} \int
\frac{d^3Q}{(2\pi)^3}\, e^{i{\bf Q}\cdot{\bf r}} \big(H_{{\bm
q}_+} \rho_{{\bm q}_+ {\bm q}_-} - \rho_{{\bm q}_+{\bm q}_-}
H_{{\bm q}_-}\big) = - \frac{i}{\hbar} \sum_{{\bm \kappa}} \int
\frac{d^3Q}{(2\pi)^3}\, e^{i{\bf Q}\cdot{\bf r}} \big(U_{{\bm
q}_+{\bm \kappa}} \rho_{{\bm \kappa}{\bm q}_-} - \rho_{{\bm
q}_+{\bm \kappa}} U_{{\bm \kappa}{\bm q}_-}\big).
\end{array}
\end{equation}
To obtain a transparent kinetic equation it is necessary to expand
$H_{{\bm q}_\pm}$ around the wave vector ${\bm q}$, which requires
some care. In this article we are working in a basis in which the
functions depend on the wave vector ${\bm q}$, and we require a
formulation of the kinetic equation that is manifestly gauge
covariant. This means that the kinetic equation should not acquire
additional terms if the basis functions are subjected to a ${\bm
q}$-dependent rotation. The end result of this requirement is that
ordinary ${\bm q}$-derivatives are replaced by covariant derivatives
(${\bm r}$-derivatives, denoted by $\nabla$, remain unchanged since
the basis does not have position dependence.) The covariant ${\bm
q}$-derivative is defined by $\frac{Df}{D{\bm q}} \equiv \pd{f}{{\bm
q}} - i\, [\bm{\mathcal{R}}, f] $, where the gauge connection matrix
$\bm{\mathcal{R}}_{ss'} = \dbkt{ {u_{ {\bm q}s } }}{ i \pd{u_{{\bm
q}s'}} {{\bm q}}}$. All this implies that in our derivation we must
replace the ordinary derivatives by covariant derivatives. Expanding
$H_{{\bm q}_\pm} \approx H_{\bm q} \pm \frac{{\bm Q}}{2} \cdot
\frac{DH_{{\bm q}}}{D{\bm q}}$ equation (\ref{frho}) can be written
as
\begin{equation}\label{eq:kineq}
\arraycolsep 0.3ex
\begin{array}{rl}
\displaystyle \pd{f_{\bm q}}{t} + \frac{i}{\hbar}\, [H_{{\bm q}},
f_{\bm q}] + \frac{1}{2\hbar}\, \{ \frac{DH_{{\bm q}}}{D{\bm q}}
\cdot \nabla f_{\bm q} \} = - \frac{i}{\hbar} \sum_{{\bm \kappa}}
\int \frac{d^3Q}{(2\pi)^3}\, e^{i{\bf Q}\cdot{\bf r}} \big(U_{{\bm
q}_+{\bm \kappa}} \rho_{{\bm \kappa}{\bm q}_-} - \rho_{{\bm
q}_+{\bm \kappa}} U_{{\bm \kappa}{\bm q}_-}\big),
\end{array}
\end{equation}
\end{widetext}
where $\{\}$ denotes an anti-symmetrized dot product, that is the
symmetrized scalar product between vectors ${\bm a}$ and ${\bm b}$
is given by $\{ {\bm a}\cdot{\bm b} \} \equiv {\bm a}\cdot{\bm b}
+ {\bm b}\cdot{\bm a} $. The scattering term represented by the
RHS is dealt with in the appendix.

In a constant and uniform electric field the total Hamiltonian
contains an additional term containing the electromagnetic
potential $V = e{\bm E}\cdot\hat{\bm r}$, where $\hat{\bm r}$ is
the position operator. This term is diagonal in real space.
Following the spirit of the derivation presented above, this term
appears on the right side of the kinetic equation in the same way
as $U$, and is expanded as follows
\begin{equation}
\frac{i}{\hbar} \int \frac{d^3Q}{(2\pi)^3}\, e^{i{\bm Q}\cdot{\bf
r}} \tbkt{{\bm q}_+} {[V, \rho]}{{\bm q}_-} \approx
-\frac{1}{\hbar}\, {\bm \nabla} V \cdot \frac{Df_{{\bm q}}}{D{\bm
q}}.
\end{equation}
The spatial gradient of the external electrical potential is equal
to the electric field $-{\nabla V} = {\bm E}$. In this work we
will be studying the response of the system to linear order in the
electric field.

When formulating a kinetic equation, which takes into account the
variation of the Wigner function in real space as well as in
momentum space, it is necessary to single out the length and
wave-vector scales relevant to the problem under study. In the work
at hand we consider carriers which are delocalized in real space and
are described by Bloch states, for which the wave vector is a good
quantum number. Nevertheless it must be borne in mind that the
carrier occupies a finite range of real and momentum space, denoted
by $\Delta{\bm r}$ and $\Delta{\bm q}$ respectively, which are
determined in such a way as to be consistent with the Heisenberg
uncertainty principle. In the course of a scattering event in which
a carrier with wave vector ${\bm q}$ interacts with the potential of
an impurity and its wave vector changes from ${\bm q}$ to ${\bm
\kappa}$, it is necessary as well as physical to assume that the
wave-vector spread $\Delta{\bm q}$ associated with the carrier size
is m! uch smaller than the typical momentum transfer in scattering
processes ${\bm \kappa} - {\bm q}$. Furthermore, it is assumed that
the magnetization ${\bm M}$ varies over length scales much larger
than interatomic separations. With these assumptions, the kinetic
equation in an electric field ${\bm E}$ takes the form (in agreement
with the form found by Carruthers and Zachariasen \cite{car83})
\begin{widetext}
\begin{equation}\label{eq:equikineq}
\pd{f}{t} + \frac{i}{\hbar}\, [H_v + H_{pd}, f]+ \frac{1}{2\hbar}
\left\{ \frac{D}{D{\bm q}}\, (H_v + H_{pd}) \cdot \nabla f
\right\} - \frac{1}{2\hbar} \left\{ \nabla H_{pd} \cdot
\frac{Df}{D{\bm q}} \right\} + \hat J (f) = \Sigma_E,
\end{equation}
\end{widetext}
where $\Sigma_E = -e{\bm E}\cdot \big(Df/D{\bm q}\big)$ is the
covariant form of the usual source term due to ${\bm E}$. The term
$\hat{J}(f)$ represents the scattering term, which is discussed in
detail in the appendix. The scattering term takes into account the
effect of the potential $\mathcal{U}$, which represents the part
of the Hamiltonian $H_{\rm Mn}$ which is off-diagonal in wave
vector. An explicit form for the scattering term will be given
below when we discuss the solution of the kinetic equation in an
electric field.

\section{Solution of the kinetic equation}

The equilibrium distribution $f_\mathrm{eq}$ is the solution to Eq.\
(\ref{eq:equikineq}) in the absence of external fields, $\Sigma_E =
0$. To leading order in $|U|^2$ this solution is $f_\mathrm{eq} =
f_0 (H_v + H_{pd})$, with $f_0$ the Fermi-Dirac function. It is
straightforward to check that this form of the Wigner function
satisfies the kinetic equation (\ref{eq:equikineq}) when the RHS is
equal to zero. The form of his solution shows that in equilibrium
the spin polarization of the holes follows the magnetization of the
Mn, which is contained in the exchange part of the Hamiltonian
$H_{pd}$.

Next, in the linear response regime, we search for a solution of the
kinetic equation for nonzero $\Sigma_E$, which will yield the spin
density induced by ${\bm E}$. Since the spin density induced by the
electric field will be a function of position and will in general
not be parallel to the local magnetization, this will immediately
give the spin torque exerted by the conduction holes on the
magnetization. The method we use to solve the kinetic equation is as
follows. First, we divide every matrix $\mathcal{M}$ in the problem
into $\mathcal{M}^{in} + \mathcal{M}^{out}$, where
$\mathcal{M}^{in}$ has elements only within the HH and LH subspaces,
while $\mathcal{M}^{out}$ has matrix elements only between these
subspaces. Schematically this can be summarized by
\begin{equation}
\arraycolsep 0.3ex
\begin{array}{rl}
\displaystyle M = & \displaystyle\begin{pmatrix} in & out \cr out
& in
\end{pmatrix}.
\end{array}
\end{equation}
One compelling advantage of this decomposition is that commutators
and anticommutators of matrices belonging to either the \textit{in}
or \textit{out} sectors do not mix these sectors. The following list
covers all the possible combinations of commutators and
anticommutators of matrices belonging to either the $in$ or the
$out$ sectors
\begin{equation}
\arraycolsep 0.3ex
\begin{array}{rl}
\displaystyle [in, out] = & \displaystyle out \\ [3ex]
\displaystyle [in, in] = & \displaystyle in \\ [3ex] \displaystyle
[ out, out ] = & \displaystyle out \\ [3ex] \displaystyle \{ in,
out \} = & \displaystyle out \\ [3ex] \displaystyle \{ in, in \} =
& \displaystyle in \\ [3ex] \displaystyle \{ out, out \} = &
\displaystyle in.
\end{array}
\end{equation}
Another advantage of this decomposition is that it aids us in
constructing a physical picture of spin torques and their relation
to spin precession. The decomposition into an $in$ and an $out$
sector in effect singles out spin precession as a result of the
spin-orbit interaction. The \textit{in} sector of the density matrix
represents spins that are stationary under the action of the
spin-orbit interaction, or alternatively the fraction of the spins
that are in eigenstates of $H_v$. The \textit{out} sector on the
other hand represents spins that precess under the action of the
spin-orbit interaction. This decomposition determines which spin
torques are due to the hole spin precession, which, unlike the
precession of spin-1/2 electrons, cannot be attributed to an
effective magnetic field. \cite{dim06} Being in the weak momentum
scattering regime $\varepsilon_F\tau/\hbar \gg 1$, we do not
consider scattering in the \textit{out} sector or between the
\textit{in} and \textit{out} sectors (it can be shown that both of
these terms yield corrections linear in $|U|^2$). The Wigner
function $f$ has two parts, $f^{in}$ in the $in$ sector and
$f^{out}$ in the $out$ sector, and the kinetic equation is broken
down into two coupled equations for $f^{in}$ and $f^{out}$
\begin{subequations}
  \label{eq:inout}
  \begin{eqnarray}
    \pd{f^{in}}{t} + \frac{i}{\hbar}\, [H^{in}, f^{in}] +
    \hat J(f^{in}) & = & \Sigma_E^{in} + \Sigma_\mathrm{gr}^{in} \\
      \pd{f^{out}}{t} + \frac{i}{\hbar}\, [H^{in}, f^{out}] & = &
      \Sigma_E^{out} + \Sigma_\mathrm{gr}^{out}.
      \hspace{1em}
  \end{eqnarray}
\end{subequations}
There are two source terms in each equation, namely $\Sigma_E^{in}$
and $\Sigma_{gr}^{in}$ in the $in$ sector, and $\Sigma_E^{out}$ and
$\Sigma_{gr}^{out}$ in the out sector. To obtain these source terms
one needs to expand all quantities in the gradient of the
magnetization and keep terms to zeroth and first order in this
gradient. To zeroth order in the gradient of the magnetization the
source terms are $\Sigma_E^{in/out} = (e{\bm E}/{\hbar}) \cdot
\big(Df_\mathrm{eq}/D{\bm q} \big)^{in/out}$, which are found simply
by taking $\Sigma_E$ defined above and substituting $f_{eq}$ for the
Wigner function. When the expansion is continued to the next order,
the source terms linear in the gradient (gr) of the magnetization
are
\begin{widetext}
\begin{subequations}\label{eq:source}
  \begin{eqnarray}
    \Sigma^{in}_\mathrm{gr} & = & \frac{1}{2\hbar}
    \left\{ \nabla H^{out} \cdot \frac{Df^{out}}{D{\bm q}}\right\} -
    \frac{1}{2\hbar} \left\{ \frac{DH^{out}}{{D{\bm q}}} \cdot \nabla
    f^{out}\right\} - \frac{1}{2\hbar}
    \left\{ \frac{DH^{in}}{{D{\bm q}}} \cdot
    \nabla f^{in}\right\}
    \\
    \Sigma^{out}_\mathrm{gr} & = & \frac{1}{2\hbar} \left\{ \nabla
    H^{in}\cdot \frac{Df^{out}}{D{\bm q}} \right\} - \frac{1}{2\hbar}
    \left\{ \frac{DH^{in}}{D{\bm q}} \cdot \nabla f^{out} \right\} -
    \frac{i}{\hbar} \, \big[H^{out}, f^{out} + f^{in} \big] .
  \end{eqnarray}
\end{subequations}
To obtain Eq.\ (\ref{eq:source}) we have assumed a small spatial
gradient $\nabla H = \nabla H_{pd}$ implying a small variation
$\delta M \ll |\langle\bm{M}\rangle|$, and we worked, as stated,
to first order in $m|\bm{M}| /(2\bar{\gamma} \hbar^2k_F^2)$. After
some simplification we obtain for the scattering term acting on
$f^{in}$
\begin{equation}
\hat J (f^{in}) = \frac{f^{in} - \overline{f^{in}}}{\tau} +
\frac{\overline{\Gamma_s^2}}{\tau} \, f^{in} - \frac{1}{\tau}\,
\overline{\Gamma_s \, f^{in} \, \Gamma_s} + \frac{m^*}{\tau q^2
\hbar^2} \int \frac{d\Omega'}{4\pi} \left[(f^{in} - f^{in'}) \,
(H_{pd} - H_{pd}') - q\, \pd{f^{in'}}{q} \, (H_{pd} - H_{pd}')
\right],
\end{equation}
\end{widetext}
where the bar is an average over directions in momentum space,
$\tau^{-1} = N_\mathrm{Mn}|U|^2 m^*q/ (V \pi \hbar^3)$, $f \equiv f
(q, \theta, \phi)$ and $f' \equiv f (q, \theta', \phi')$, $\theta$
and $\phi$ are the polar and azimuthal angles of ${\bm q}$
(analogously for ${\bm q}'$), and $m^\ast$ is the carrier effective
mass, which is $m/(\gamma_1 - 2\bar\gamma)$ in the HH subspace, and
$m/(\gamma_1 + 2\bar\gamma)$ in the LH subspace.

For simplicity and without loss of generality we choose ${\bm E}
\parallel \hat{\bm y}$ and $\langle\bm{M}\rangle \parallel \hat{\bm
z}$ so that $M_{x,y} ({\bm r}) \ll M_z({\bm r})$. We solve Eqs.\
(\ref{eq:inout}) as follows: the equation for $f^{out}$ is first
solved with $\Sigma_E^{out}$ as the initial source, and the solution
$f_E^{out}$ thus obtained is substituted into
$\Sigma_\mathrm{gr}^{out}$ and $\Sigma_\mathrm{gr}^{in}$. The
equation for $f^{in}$ is solved in an analogous fashion. The
solutions to the equations for $f^{in}$ and $f^{out}$ involve
expressions of the form $e^{iH^{in}t/\hbar} \mathcal{M}
e^{-iH^{in}t/\hbar}$, and $f^{out}$ is easily found. This is because
in the \textit{out} sector the product $e^{iH^{in}t/\hbar}
\mathcal{M}^{out} e^{-iH^{in}t/\hbar}$ contains only functions of
time of the form $\sin \omega t$ and $\cos\omega t$, with $\omega =
2\hbar \bar\gamma q^2/m$ the energy difference between the HH and LH
bands when the magnetization is zero. The steady-state solution for
$f^{out}$ therefore involves only a straightforward time integral of
the kind customarily encountered in linear-response theories. The
equation for $f^{in}$ takes more effort due to the presence of the
scattering term and we only summarize the method here (it is
described in detail in a recent publication by two of us\
\cite{steadystate}). The $in$ sector represents the part of the
Wigner function that is stationary under the action of $H_v$.
Nevertheless, the full Hamiltonian is $H_v + H_{pd}$, and the
commutator $[H^{in}_{pd}, f^{in}]$ is not zero. In a manner similar
to the decomposition of $f$ into $f^{in}$ and $f^{out}$, $f^{in}$
itself is split into a part that commutes with $H_{pd}^{in}$, and a
part that does not. It can be shown \cite{steadystate} that the
commuting part yields a correction to the Wigner function that is
linear in $\tau$ while the non-commuting part gives a correction
that does not depend on $\tau$. However, we find that all
contributions to $f$ average to zero over directions in momentum
space except $f^{out}_\mathrm{gr}$. This implies that all
contributions from $f^{in}$ average to zero over directions in
momentum space. $f^{out}_\mathrm{gr}$ gives rise to a spin density
$\bm{\mathcal{S}}$ that is independent of scattering. It is
discussed in detail below.

\section{Spin torques}

The only contribution to the spin density in an electric field
comes from $f^{out}_\mathrm{gr}$. The three components of the spin
density $\mathcal{S}$ that this correction to the Wigner function
yields are
\begin{subequations}
  \label{eq:torque}
  \begin{eqnarray}
    \mathcal{S}_x & =  &
      \frac{eE_ym^{1/2}}{\varepsilon_F^{3/2}} \, \bigg( \eta_x \pd{M_x}{y}
      - \zeta_x\pd{M_y}{x} \bigg) \\
      \mathcal{S}_y & = & \frac{eE_ym^{1/2}}{\varepsilon_F^{3/2}}
      \, \bigg( \eta_y \pd{M_y}{y} - \zeta_y \pd{M_x}{x} \bigg) \\
      \mathcal{S}_z & = & \frac{eE_ym^{1/2}}{\varepsilon_F^{3/2}}\,
      \eta_z \pd{M_z}{y} .
  \end{eqnarray}
\end{subequations}
These equations are the central result of our work. The
dimensionless quantities $\eta_i$ and $\zeta_i$, with $i = x,y,z$
are functions of the Luttinger parameters $\gamma_1$ and
$\bar\gamma$. For GaMnAs we find (all $\times 10^{-4}$) $\eta_x =
\eta_z = 3.66$, $\eta_y = 5.52$, $\zeta_x = 11.56$ and $\zeta_y =
6.16$. The steady-state spin density is not collinear with the
magnetization, so there will be a torque on the magnetization giving
a precession frequency of magnitude $J_{pd}|{\bm {\mathcal S}}|$.
Taking $p = 1.2 \times 10^{20}$~cm$^{-3}$, $E_y = 100$~kV/m, and
estimating the change in the magnetization as 20$\%$ over one
lattice spacing, the time scale of this precession is 200~ns -- less
than in metals, but ${\bm M}$ itself is also typically one order of
magnitude smaller.

\subsection{Discussion}

The fact that the spin torque comes only from $f^{out}_\mathrm{gr}$
implies that the steady-state spin density is due to precession
under the action of both the spin-orbit interaction and the exchange
field. The fraction of the spins that is conserved, which would
yield a term $\propto \tau$, gives a contribution that averages to
zero in momentum space. The quantities $\eta_i$ and $\zeta_i$
decrease with increasing spin-orbit interaction (given by
$\bar\gamma$), suggesting the spin-orbit interaction reduces the
spin torque. This agrees with the finding that there is no
electrically-induced spin density in the corresponding nonmagnetic
systems, \cite{ivc78} i.e. in the limit of large spin-orbit
interaction $\bar{\gamma}$. This limit is equivalent to restoring
the spherical symmetry of the Luttinger Hamiltonian of Eq.\
(\ref{eq:HLutt}), which in ferromagnetic semiconductors is broken by
the magnetization.

An important difference from ferromagnetic metals is that, in Eq.\
(\ref{eq:torque}), there is no contribution from scattering, either
scalar or spin-dependent. This fact indicates that the dominant spin
torque in ferromagnetic semiconductors in the weak momentum
scattering limit is intrinsic. This observation agrees with the
results of Jungwirth \textit{et al}., \cite{jun02} who studied the
anomalous Hall effect in ferromagnetic semiconductors in the regime
$\varepsilon_F \tau/\hbar \gg 1$ and found similarly that the role
of scattering is secondary. It is also related to the absence of
electrically-induced spin polarization in bulk nonmagnetic
zincblende materials. Generally, such a spin polarization is due to
the fraction of spins that is conserved \cite{steadystate} and is
linear in $\tau$, but this spin polarization is forbidden by
symmetry in zincblende lattices. \cite{ivc78} The magnetization
breaks the cubic symmetry of the lattice and gives a steady-state
spin density, but the term linear in $\tau$ still averages to zero.
We come back to the comparison of our result with result found for
ferromagnetic metals in the next section.

We find that an electric field ${\bm E} \parallel \hat{\bm x}$
corresponds to the permutation $x \leftrightarrow y$ in Eq.\
(\ref{eq:torque}). Yet for a given orientation of ${\bm E}$, unlike
in ferromagnetic metals, in ferromagnetic semiconductors there is no
symmetry between the different components of the spin density for
the following reason. In metals spin is conserved and spin torques
can be derived phenomenologically directly from the
Landau-Lifshitz-Gilbert equation (the so-called
\textit{book-keeping} argument \cite{li04, dui07}). One assumes an
itinerant spin passes a localized moment at ${\bm r}$, lines up with
it, then moves on to another moment at ${\bm r} + \delta{\bm r}$ and
exerts a torque on this moment. This relates ${\bm M}({\bm r} +
\delta{\bm r})$ to ${\bm M}({\bm r})$ and gives a simple
vector-product form for ${\bm S}({\bm r})$. \cite{dui07, pie07,
koh06, tse06} In ferromagnetic semiconductors the spin-orbit
interaction acts to randomize the itinerant spin moving between
${\bm r}$ and ${\bm r} + \delta{\bm r}$, and there is no simple
relationship between ${\bm M}({\bm r} + \delta{\bm r})$ and ${\bm
M}({\bm r})$. Such a book-keeping argument is thus not valid and
there is no symmetry in the final expression for the spin density.

We would like to comment on one last aspect of the relationship
between the hole spin polarization and the magnetization in
ferromagnetic semiconductors. The calculation presented in this work
relies on a mean-field description of the magnetization and hole
spin polarization. In this picture the itinerant holes are subject
to an average magnetic field due to the Mn$^{2+}$ ions, and the
Mn$^{2+}$ ions in turn are subject to the itinerant hole spin
polarization, which can also be regarded as an average magnetic
field.\cite{jun06} Since it is assumed that the spin-orbit
interaction has spherical symmetry, there is no easy axis for the
magnetization in the absence of an electric field. However, once the
electric field is applied it is natural to ask whether the direction
of the electric field provides an easy axis for the magnetization,
in other words whether the magnetization in the direction of the
electric field increases. We find that this indeed is true, but the
increase in the magnetization is second order in the ratio
$H_{pd}/\varepsilon_F$ and is not significant.

In ferromagnetic metals, in which spin-orbit coupling is negligible,
angular momentum is conserved. As a result spin torques in these
materials can be encapsulated into a set of simple, compact,
rotationally-invariant vectorial expressions. In ferromagnetic
semiconductors, in which spin-orbit interactions are usually the
dominant energy scale, angular momentum is not conserved and the
final expressions for the spin torques cannot be expected to have
rotational invariance. In principle spin-orbit interactions, which
couple the spin and the lattice, should give magnetic anisotropy and
anistropic spin torques as well. The anisotropy in our result for
the spin density is thus a direct result of the intrinsic spin-orbit
interactions.

\subsection{Parameters and applicability for GaMnAs}

We shall assume a doping density $n_{Mn} = p = 1.2 \times 10^{20}cm^{-3}$, corresponding to $x=2.2\%$, $J_{pd} = 54$ meV
nm$^{3}$ as discussed in Ref.~[\onlinecite{jun06}] and the lattice constant $a = 5.6533 \AA$. The Fermi energy is found as
\begin{equation}
\arraycolsep 0.3ex
\begin{array}{rl}
\displaystyle \big(\frac{2m_h\varepsilon_F}{\hbar^2}\big)^{3/2} +
\big(\frac{2m_l\varepsilon_F}{\hbar^2}\big)^{3/2} = 3\pi^2 n \\
[3ex] \displaystyle \varepsilon_F = 1.633 \,
\frac{\hbar^2}{2m_0}\, (3\pi^2n)^{2/3} = 2.1 \times 10^{-20}\,
{\rm J}
\end{array}
\end{equation}
and the heavy and light hole Fermi wave vectors are $k_h = 1.43
\times 10^9 {\rm m}^{-1}$ and $k_l = 0.55 \times 10^9 {\rm m}^{-1}$.
The heavy hole and light hole masses are $m_h = 0.538 \times
10^{-30}$kg and $m_l = 0.076 \times 10^{-30}$kg. These numbers also
give the magnitude of the effective field $|H_{pd}| = n_{Mn}\,
J_{pd}\, <S_{Mn}> = 2.52 \times 10^{-21}$ J, meaning that the ratio
$|H_{pd}|/\varepsilon_F = 0.12$, so it is safe to do perturbation
theory.

We also want to work out $\varepsilon_F\tau_{p}/\hbar$. The Fermi
energy is $2.1\times 10^{-20}$J, which means
$\varepsilon_F\tau_p/\hbar > 1$ for any momentum scattering time
$\tau_p \ge 5 \times 10^{-15}$s. For example for
$\varepsilon_F\tau_p/\hbar \approx 10 $ we require $\tau_p = 5
\times 10^{-14}$s, corresponding to a light-hole mobility of
approximately $1000$cm$^2$/Vs and a heavy-hole mobility of
approximately $200$cm$^2$/Vs. Thus the theory is on very firm ground
even for extremely low mobilities.

\section{Domain-wall motion} \label{sec:dws}

As an application of the central result in Eq.~(\ref{eq:torque}) we
calculate the spin torque on a domain wall, and the resulting
domain-wall velocity. We choose the current and variation of
magnetization in the $y$-direction. Furthermore, we use
$\eta_x=\eta_z$ such that Eq.~(\ref{eq:torque}) reduces to
\begin{equation}\label{eq:DomWall:spindensity}
{\bm {\mathcal S}}=\frac{e E_y m^{1/2}\eta_x}{\varepsilon_{\rm
F}^{3/2}} \frac{\partial}{\partial y}\left[\mathbf{M} +
\left(\frac{\eta_y}{\eta_x}-1\right)M_{y}\hat{\mathbf{y}}\right],
\end{equation}
with $\hat{\mathbf{y}}$ a unit vector in the $y$-direction. The
spin-transfer torque that acts on the magnetization is given by
\begin{equation}\label{eq:DomWall:precession}
\left.\frac{\partial\mathbf{M}}{\partial t}\right|_{\rm current
}\!=-\frac{J_{\rm pd}}{\hbar^2}
\mathbf{M}\times\boldsymbol{\mathcal{S}}.
\end{equation}
Using Eq.~(\ref{eq:torque}) we rewrite this as an equation for a
unit vector $\bm{\Omega}$ in the direction of magnetization, i.e.,
${\bf M} = n_{\rm Mn} J_{\rm pd} S_{\rm Mn} \bm{\Omega}$, with
$S_{\rm Mn}=5/2$ the spin of one Mn atom. We find that
\begin{equation}\label{eq:DomWall:unitvector}
\left.\frac{\partial\mathbf{\Omega}}{\partial t}\right|_{\rm
current }\!=-v \bm{\Omega} \times \frac{\partial}{\partial y}
\left[\bm{\Omega}+\left(\frac{\eta_y}{\eta_x}-1\right)
\Omega_{y}\hat{\mathbf{y}}\right],
\end{equation}
with the velocity $v$ given by
\begin{equation}
\label{eq:stvelocity}
  v = \frac{n_{\rm
  Mn} e E m^{1/2} \eta_x J_{\rm pd}^2 S_{\rm Mn}}{\hbar^2 \epsilon_{\rm
  F}^{3/2}}~.
\end{equation}
The result for the current-induced torques in
Eq.~(\ref{eq:DomWall:unitvector}) has the form of an anisotropic
dissipative spin transfer torque. \cite{dui07} The reactive spin
transfer torque contribution is equal to zero. These results are
understood by noting that we have considered strong spin-orbit
interactions, and that have done perturbation theory in the
magnetization.

It is common to define a dissipative coefficient \cite{koh06,dui07}
$\beta$ such that $v \sim \beta j$, with $j$ the current density.
Because our result for the spin transfer torque is independent of
$\tau$, and because $j \sim \tau$, we would find that $\beta \sim
1/\tau$, i.e., resistivity-like. This is somewhat surprising as
recent studies \cite{garate2008} indicate that the Gilbert damping
constant $\alpha_G$, which is believed to be similar though not
exactly equal to $\beta$, predominantly has intra-band contributions
that are conductivity-like. However, a direct comparison is not
possible because in the present paper we perform an expansion in the
magnitude of the magnetization whereas Ref.~[\onlinecite{garate2008}]
calculates $\alpha_G$ by determining the transverse response
function.

The velocity $v$ divided by the Gilbert damping constant provides an
estimate for the domain wall velocity $\dot X$, \cite{li04} so that
\begin{equation}
  \dot X \sim \frac{v}{\alpha_G}~.
\end{equation}
Although Sinova {\it et al.} do not explicitly consider the regime
of parameters quoted in the previous section, their calculations
\cite{sinova2004} (see also Ref.~[\onlinecite{garate2008}]) suggest
that the Gilbert damping is very small $\alpha_G \sim 0.001$ in this
regime. Using this result we find that $\dot X \sim 1$ m/s, in
agreement with experimental results for the domain-wall velocity.
\cite{yamanouchi06}

To investigate more quantitatively the effect of the anisotropy in
the spin transfer torque, determined by the ratio $\eta_y/\eta_x$,
we consider specific model for a magnetic domain wall. We consider a
thin film, in which there is a constant hard-axis anisotropy
$K_\perp$ perpendicular to the film and an easy-axis anisotropy
$K_{\rm E}$. Within the model for a rigid domain wall proposed by
Tatara and Kohno \cite{tatara04} (see also
Ref.~[\onlinecite{duine2007}]), the domain wall is described by two
collective coordinates: the position $X(t)$ and the chirality
$\phi_0(t)$. The chirality is the angle with which the magnetic
moment in the center of the domain wall tilts out of the easy plane.
Using the results from Ref.~[\onlinecite{tatara04}] and
Ref.~[\onlinecite{duine2007}] we find the equations of motion for the
domain-wall collective coordinates. They are given by
\begin{subequations}\label{eq:DomWall:EOM}
  \begin{eqnarray}\label{eq:DomWall:EOM1}
  \frac{\dot{X}}{\lambda}-\alpha_G\dot{\phi}_0 & =
  & \frac{K_\perp}{\hbar}\sin2\phi_0 \\\label{eq:DomWall:EOM2}
    \dot{\phi}_0 +\alpha_G \frac{\dot{X}}{\lambda} &=&
   \frac{v}{\lambda}
\left(1+\delta
      \cos^2\phi_0\right),
  \end{eqnarray}
\end{subequations}
where $\lambda=\sqrt{J/K_{\rm E}}$ is the width of the domain wall.
Note that $\delta$ goes to zero for $\eta_x\rightarrow\eta_y$. Note
that, in addition usual dissipative spin transfer torque
contribution to these equations discussed in earlier work,
\cite{tatara04,duine2007} we find a chirality-dependent anisotropic
contribution proportional to $\delta$.

\begin{figure}[h!]
\centering
\includegraphics[width=8cm]{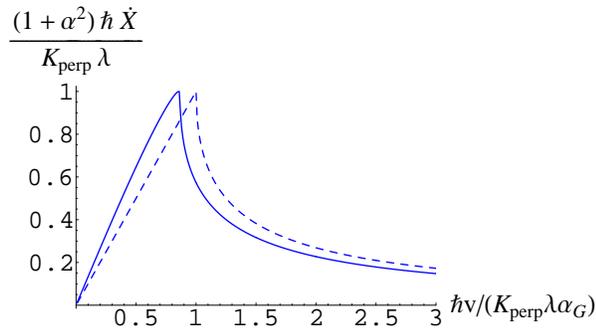}
\caption{Average drift velocity as a function of the applied
electric field and for $\delta=1/3$ (solid line) and $\delta=0$
(dashed line).}\label{fig:driftvelocity}
\end{figure}

The above equation can be solved analytically. From this we obtain
the average drift velocity as a function of the applied electric
field, as shown in Fig.~\ref{fig:driftvelocity}.
From this figure, we observe a Walker-breakdown-like behavior,
\cite{schryer1974} i.e., the domain wall velocity reaches a maximum
and then becomes smaller. Physically, the breakdown is due to the
transition of rigid motion to precessional motion of the domain wall
and is well-known from field-driven domain-wall motion. Our results
are understood from the fact that dissipative spin-transfer torque
enters the equations of motion for the domain wall in the same way
as an external magnetic field. Note that the anisotropy $\delta$
alters the result for the domain-wall velocity somewhat with respect
to the isotropic ($\delta=0$) situation, but plays no qualitatively
important role.

As a final remark, we note that in the calculations presented here
we have neglected the effects of finite temperature \cite{duine2007}
and pinning of the domain wall. This, in addition to the fact that
the experiments of Yamanouchi {\it et al.} \cite{yamanouchi06} are
in a different regime of doping than considered here, makes a direct
quantitative comparison not possible.

\section{Conclusions and Acknowledgements} 
\label{sec:concl} 

In conclusion, we have established a microscopic theory of spin
transfer in III-V ferromagnetic semiconductors for the case of
strong spin-orbit coupling. We have applied our results to the case
of current-driven domain wall motion and have estimated the
resulting domain-wall velocities. We find domain-wall velocities
that are of the same order of magnitude as experiments, although the
available experimental results \cite{yamanouchi06} are in a
different regime of parameters than considered in this paper.
Therefore, a more quantitative comparison between theory and
experiment is at present not feasible.

We wish to acknowledge enlightening discussions with A.~H.\
MacDonald, M.~Tsoi, G.~E.~W.\ Bauer, L.~W.~Molenkamp, G.~Schmidt,
B.~J.~van~Wees, M.~van Veenendaal, D.~J.\ Keavney and Zhuge Liang.
The research at Argonne National Laboratory was supported by the US
Department of Energy, Office of Science, Office of Basic Energy
Sciences, Contract No.\ DE-AC02-06CH11357. The research at Utrecht
University was supported by the Netherlands Organization for
Scientific Research (NWO) and by the European Research Council (ERC)
under the Seventh Framework Program (FP7).

\appendix

\section{Scattering term}

The scattering term $\hat J (f)$ is
\begin{equation}
\hat J(f) \approx \frac{N_\mathrm{Mn}}{\hbar^2} \lim_{\eta
\rightarrow 0} \int_{0}^\infty \! dt'\, e^{- \eta t'} \left[\hat
U, e^{- i \hat H t'} [\hat U, \hat f]\, e^{ i \hat H t'}
\right]_{{\bm q}{\bm q}},
\end{equation}
with $\eta$ a regularization factor and the impurity average
$\bkt{H_\mathrm{Mn}^{{\bm q} {\bm q}'}H_\mathrm{Mn}^{{\bm q}' {\bm
q}}}_{{\bm q}\ne {\bm q}'} = N_\mathrm{Mn}\, |U|^2 \, \big(\openone
+ \Gamma_s \big)$, where $N_{Mn}$ is the \textit{number} of Mn
impurities and $\Gamma_s = J_{pd}^2\, ({\bm s}\cdot{\bm S})^2/(|U|^2
V^2)$. The derivation of this general form of the scattering term is
discussed in a recent paper \cite{dim06} and the notation will be
explained in detail below. In terms of the Wigner distribution the
scattering term can be expressed as
\begin{widetext}
\begin{equation}
\arraycolsep 0.3ex
\begin{array}{rl}
\displaystyle \hat J(f_{\bm q}) \approx & \displaystyle
\frac{n_{Mn}}{\hbar^2} \int \frac{d^3\kappa}{(2\pi)^3} \,
\int^{\infty}_0 dt'\, e^{- \eta t'}\, U_{{\bm q}{\bm \kappa}} e^{-
i
H_{{\bm \kappa}} t'}\big(U_{{\bm \kappa}{\bm q}} f_{\bm q} - f_{\bm \kappa} U_{{\bm \kappa}{\bm q}} \big) \, e^{i H_{{\bm q}} t'} \\
[3ex] & \displaystyle - \frac{n_{Mn}}{\hbar^2} \int
\frac{d^3\kappa}{(2\pi)^3} \, \int^{\infty}_0 dt'\, e^{- \eta
t'}\, e^{- i H_{{\bm q}} t'} \big( U_{{\bm q}{\bm \kappa}} f_{\bm
\kappa} - f_{\bm q} U_{{\bm q}{\bm \kappa}} \big) \, e^{i H_{{\bm
\kappa}} t'} U_{{\bm \kappa}{\bm q}} .
\end{array}
\end{equation}
\end{widetext}
We must note that, in the approximation we are using, the scattering
term acts only on $f^{in}$, which brings about some simplifications.
These become apparent if we look at the explicit form of this term
and note that, because it involves only $f^{in}$, this term commutes
with the time evolution operators.
\begin{widetext}
\begin{equation}
\arraycolsep 0.3ex
\begin{array}{rl}
\displaystyle \hat J(f^{in}) = & \displaystyle
\frac{n_{Mn}}{\hbar^2} \int \frac{d^3\kappa}{(2\pi)^3} \,
\int^{\infty}_0 dt'\, e^{- \eta t'}\, U \big( e^{- i H_{{\bm
\kappa}} t'} U \, e^{i H_{{\bm q}} t'} f^{in}_{\bm q} -
f^{in}_{\bm \kappa} e^{- i H_{{\bm \kappa}} t'}U \, e^{i H_{{\bm
q}} t'} \big) \\ [3ex] - & \displaystyle \frac{n_{Mn}}{\hbar^2}
\int \frac{d^3\kappa}{(2\pi)^3} \, \int^{\infty}_0 dt'\, e^{- \eta
t'}\, \big( e^{- i H_{{\bm q}} t'}\, U \, e^{i H_{{\bm \kappa}}
t'} f^{in}_{\bm \kappa} - f^{in}_{\bm q} e^{- i H_{{\bm q}} t'}U
\, e^{i H_{{\bm \kappa}} t'} \big) U .
\end{array}
\end{equation}
In the approximation used in this paper, the Hamiltonian entering
the scattering term is the projected $2\times 2$ Hamiltonian for
each subspace. The scattering potential has two parts, one a scalar
and one which is spin-dependent. Taking into account also the
exchange splitting of the bands, there are three contributions to
the scattering term: scalar potential + kinetic energy ($\equiv
\hat{J}_0$), spin-dependent potential + kinetic energy ($\equiv
\hat{J}_m$), scalar potential + exchange energy ($\equiv
\hat{J}_s$). The former two sum up to
\begin{equation}
\arraycolsep 0.3ex
\begin{array}{rl}
\displaystyle \hat J_0 (f^{in}) + \hat J_m (f^{in}) \approx &
\displaystyle \frac{2\pi n_{Mn}}{\hbar} \int
\frac{d^3\kappa}{(2\pi)^3} \, \big(
\frac{1}{2}\{ U^2, f^{in}_{\bm q} \} - U f^{in}_{\bm \kappa} U \big) \, \delta(\frac{\hbar^2\kappa^2}{2m^*} - \frac{\hbar^2q^2}{2m^*}) \\
[3ex] = & \displaystyle \frac{n_{Mn} m^*q}{4\pi^2\hbar^3} \int
d\Omega' \, \big( \frac{1}{2}\{ U^2, f^{in}_{\bm q} \} - U
f^{in'}_{\bm q} U \big) = \frac{n_{Mn} m^*q}{\pi \hbar^3} \, \big(
\frac{1}{2}\{ \overline{U^2}, f^{in}_{\bm q} \} -
\overline{Uf^{in}_{\bm q}U} \big) \\ [3ex] =
& \displaystyle \frac{1}{\tau}\, \big( \frac{1}{2}\{ \overline{\Gamma^2}, f^{in}_{\bm q} \} - \overline{\Gamma f^{in}_{\bm q} \Gamma} \big) \\
[3ex] \displaystyle \frac{1}{\tau} = & \displaystyle \frac{n_{Mn}
|U|^2 m^*q}{\pi \hbar^3} \,\,\, {\rm and} \,\,\, U = |U| \Gamma
\\ [3ex]
\displaystyle \overline{f^{in}_{\bm q}} = & \displaystyle
\frac{1}{4\pi} \int d\Omega' \, f^{in'}_{\bm q}.
\end{array}
\end{equation}
We have used the notation $f^{in'}_{\bm q} \equiv f^{in}(q,
\Omega')$. $|U|^2$ is a scalar and $\Gamma$ is a dimensionless
matrix, which is written as $\Gamma = \openone + \Gamma_s$, with
$\Gamma_s = 1/2 \, {\bm \Gamma}_s\cdot{\bm \sigma}$. Notice that
$\Gamma_s$ has angular dependence because we are in the basis of
eigenstates of the Luttinger Hamiltonian. Thus these two
contributions to the scattering term can be rewritten as
\begin{equation}
\arraycolsep 0.3ex
\begin{array}{rl}
\displaystyle \hat J_0 (f) + \hat J_m (f) = & \displaystyle
\frac{1}{\tau}\, \big[ \frac{1}{2}\{ \overline{(\openone +
\Gamma_s)^2} , f \} - \overline{(\openone + \Gamma_s)\, f
(\openone + \Gamma_s)}\big] = \frac{1}{\tau}\, \big[ f + \{
\overline{(\Gamma_s + \frac{\Gamma_s^2}{2})}, f \} - (\overline{f}
+
\overline{\{ \Gamma_s, f \}} + \overline{\Gamma_s f \Gamma_s}) \big] \\
[3ex] = & \displaystyle \frac{f - \overline{f}}{\tau} +
\frac{1}{\tau} \, \{ \overline{\Gamma_s + \frac{\Gamma_s^2}{2}}, f
\} - \frac{1}{\tau}\, \overline{ \{ \Gamma_s, f \}} -
\frac{1}{\tau}\, \overline{\Gamma_s \, f \, \Gamma_s}.
\end{array}
\end{equation}
\end{widetext}
We think of $\tau$ as a characteristic scattering time. The
explicit form of the potential, determined previously, is
\begin{equation}
\arraycolsep 0.3ex
\begin{array}{rl}
\displaystyle U^2 = & \displaystyle N_i \mathcal{U}^2 \, \big[1 +
2\alpha \, ({\bm s}\cdot{\bm S}) + \alpha^2 ({\bm s}\cdot{\bm
S})^2\big],
\end{array}
\end{equation}
where $\alpha = \mathcal{V}/\mathcal{U}$. Everything must be
averaged over the impurity configuration as well as directions in
momentum space, and then it needs to be transformed into the
eigenstate basis and projected onto LH and HH. When we do that,
the term linear in $\alpha$ above contains only $S_z$, which, when
projected onto LH and HH gives something that averages to zero
over angles. Moreover, the configuration average of $\Gamma_s^2$
gives something which, when restricted to the HH and LH subspaces,
is proportional to the identity matrix, so contributes only the
scalar part of the scattering term.
\begin{equation}
\arraycolsep 0.3ex
\begin{array}{rl}
\displaystyle \hat J_0 (f) + \hat J_m (f) = & \displaystyle
\frac{f - \overline{f}}{\tau} + \frac{\overline{\Gamma_s^2}}{\tau}
\, f - \frac{1}{\tau}\, \overline{ \{ \Gamma_s, f \}} -
\frac{1}{\tau}\, \overline{\Gamma_s \, f \, \Gamma_s}.
\end{array}
\end{equation}
We separate the action of $\hat{J}_m$ on the scalar and
spin-dependent parts $n$, and $S$, of the Wigner distribution $f =
n \openone + S$. First on $n$, which is written as $n =
\overline{n} + \nu$, where $\nu$ is the anisotropic part
\begin{equation}\label{eq:Jn}
\arraycolsep 0.3ex
\begin{array}{rl}
\displaystyle \hat J_0 (n) + \hat J_m (n) = & \displaystyle (1 +
\overline{\Gamma_s^2})\, \frac{\nu}{\tau} - \frac{2}{\tau}\,
\overline{\Gamma_s \nu} - \frac{1}{\tau}\, \overline{\Gamma_s^2 \,
\nu}
\end{array}
\end{equation}
Averaged over impurities $\Gamma_s^2$ gives
\begin{equation}
\arraycolsep 0.3ex
\begin{array}{rl}
\displaystyle \Gamma_s^2 = & \displaystyle N_i \, |\mathcal{V}|^2
[s_\perp^2 (S_x^2 + S_y^2) + s_z^2 S_z^2] = \gamma^2_{h,l}\,
\openone,
\end{array}
\end{equation}
the latter identity being valid because the matrix elements of the
$S_i^2$ restricted to the HH and LH subspaces are proportional to
the identity matrix. We also need to average $\Gamma_s\, \sigma_i
\, \Gamma_s$, for which we note that $\sigma_i \, \sigma_j \,
\sigma_i = - \sigma_j$ for $i\ne j$. Averaged over impurities
\begin{equation}
\arraycolsep 0.3ex
\begin{array}{rl}
\displaystyle \Gamma_s\, \sigma_x \, \Gamma_s = & \displaystyle
\frac{1}{4}\, (\Gamma_x^2 - \Gamma_y^2 - \Gamma_z^2) \, \sigma_x
\\ [3ex]
\displaystyle \Gamma_s\, \sigma_y \, \Gamma_s = & \displaystyle
\frac{1}{4}\, (\Gamma_y^2 - \Gamma_x^2 - \Gamma_z^2) \, \sigma_y
\\ [3ex]
\displaystyle \Gamma_s\, \sigma_z \, \Gamma_s = & \displaystyle
\frac{1}{4}\, (\Gamma_z^2 - \Gamma_x^2 - \Gamma_y^2) \, \sigma_z.
\end{array}
\end{equation}
This tells us that in the term $\overline{\Gamma_s \, S \,
\Gamma_s}$ only the average of $S$, which we shall call
$\overline{S}$, survives. Then, writing $S = \overline{S} + \Xi$
\begin{equation}\label{eq:Js}
\arraycolsep 0.3ex
\begin{array}{rl}
\displaystyle \hat J_0 (S) + \hat J_m (S) = & \displaystyle
\frac{\Xi}{\tau} + \frac{\overline{\Gamma_s^2}}{\tau} \,
(\overline{S} + \Xi) - \frac{1}{\tau}\, \overline{ \{ \Gamma_s,
\Xi \}} - \frac{1}{\tau}\, \overline{\Gamma_s \, \overline{S} \,
\Gamma_s}.
\end{array}
\end{equation}
Looking at Eq.\ (\ref{eq:Jn}) and (\ref{eq:Js}) we see that if we
ignore the term linear in $\Gamma_s$ in each of them then they do
not mix the scalar and spin distributions. We shall work for now
in this approximation, which is justified because the terms
omitted are higher order in the disorder potential. Then we can
write
\begin{equation}
\arraycolsep 0.3ex
\begin{array}{rl}
\displaystyle \hat J_0 (n) + \hat J_m (n) = & \displaystyle
\frac{\nu}{\tau_\gamma} - \frac{1}{\tau}\, \overline{\Gamma_s^2 \,
\nu} \\ [3ex] \displaystyle \frac{1}{\tau_\gamma} = &
\displaystyle \frac{1 + \overline{\Gamma_s^2}}{\tau}\\ [3ex]
\displaystyle \hat J_0 (S) + \hat J_m (S) = & \displaystyle
\frac{\Xi}{\tau} + \frac{\overline{\Gamma_s^2}}{\tau} \,
(\overline{S} + \Xi) - \frac{1}{\tau}\, \overline{\Gamma_s \,
\overline{S} \, \Gamma_s} .
\end{array}
\end{equation}
The contribution to the scattering term due to the exchange
splitting of the bands is
\begin{widetext}
\begin{equation}
\arraycolsep 0.3ex
\begin{array}{rl}
\displaystyle \hat J_s (f) = & \displaystyle \frac{m^*}{\tau q^2
\hbar^2}\, (f - \overline{f})\, H_{pd} - \frac{m^*}{\tau q
\hbar^2} \pd{\overline{f}}{q} \, H_{pd}  + \frac{m^*}{\tau q^2
\hbar^2} \int \frac{d\Omega'}{4\pi} (f' + q \pd{f'}{q}) \,
H_{pd}'.
\end{array}
\end{equation}
\end{widetext}

\end{document}